\documentclass[11pt]{article}
\pdfoutput=1
\usepackage{jcapmod}
\usepackage{booktabs}
\usepackage[english]{babel}
\usepackage{amsmath,amssymb,amsbsy,amstext, amsthm, simplewick}
\usepackage{graphicx}
\usepackage{amsfonts}
\usepackage{amssymb}
\usepackage{upgreek}
 \usepackage{exscale,relsize}
 \usepackage[makeroom]{cancel}
\usepackage{soul}
\usepackage{slashed}

\RequirePackage{color}

\usepackage{colortbl}
\definecolor{rp}{cmyk}{0.2, 1, 0.6, 0}
\definecolor{green2}{cmyk}{0, 1, 0.5, 0}
\definecolor{lightgreen}{cmyk}{0.2, 0, 0.2, 0.2}
\definecolor{lightgray}{cmyk}{0.1,0.2,0,0.1}
\definecolor{lightgray2}{cmyk}{0.4,0.4,0,0.8}
\definecolor{black}{cmyk}{1.0,1.0,1.0,1.0}

\allowdisplaybreaks[1]

\usepackage{colortbl}
\definecolor{lightgreen}{cmyk}{0.2, 0, 0.2, 0.2}
\definecolor{lightgray}{cmyk}{0.1,0.2,0,0.1}
\definecolor{lightgray2}{cmyk}{0.1,0.1,0,0.1}

\makeatletter
\newlength{\apb@width}
\newcommand{\autoparbox}[2][c]{\settowidth{\apb@width}{#2}\parbox[#1]{\apb@width}{#2}}

\makeatother

\numberwithin{equation}{section}

\def\beq{\begin{equation}}
\def\eeq{\end{equation}}

\def\bea{\begin{eqnarray}}
\def\eea{\end{eqnarray}}

\def\beq{\begin{equation}}
\def\eeq{\end{equation}}
\def\bea{\begin{eqnarray}}
\def\eea{\end{eqnarray}}

\def\cO{{\cal O}}

\def\Mp{M_{\rm pl}}

\def\H{{\cal H}}

\def\0{{\boldsymbol 0}}
\def\k{{\boldsymbol{k}}}

\def\p{{\boldsymbol{p}}}

\def\x{{\boldsymbol{x}}}
\def\y{{\boldsymbol{y}}}

\DeclareRobustCommand{\SkipTocEntry}[4]{}

\begin{document}

\begin{titlepage}

\setcounter{page}{1} \baselineskip=15.5pt \thispagestyle{empty}

\bigskip\

\vspace{1cm}
\begin{center}

{\fontsize{20}{28}\selectfont  \sffamily \bfseries Disorder in the Early Universe}

\end{center}

\vspace{0.2cm}

\begin{center}
{\fontsize{14}{30}\selectfont   Daniel Green }
\end{center}

\begin{center}
\textsl{ Canadian Institute for Theoretical Astrophysics, Toronto, ON M5S 3H8, Canada}
\
\end{center}

\vspace{1.2cm}
\hrule \vspace{0.3cm}
\noindent {\sffamily \bfseries Abstract} \\[0.1cm]
Little is known about the microscopic physics that gave rise to inflation in our universe.  There are many reasons to wonder if the underlying description requires a careful arrangement of ingredients or if inflation was the result of an essentially random process.  At a technical level, randomness in the microphysics of inflation is closely related to disorder in solids.  We develop the formalism of disorder for inflation and investigate the observational consequences of quenched disorder.  We find that a common prediction is the presence of additional noise in the power spectrum or bispectrum.  At a phenomenological level, these results can be recast in terms of a modulating field, allowing us to write the quadratic maximum likelihood estimator for this noise.  Preliminary constraints on disorder can be derived from existing analyses but significant improvements should be possible with a dedicated treatment.  
\vskip 10pt
\hrule

\vspace{0.6cm}
 \end{titlepage}

 \tableofcontents

\newpage

\section{Introduction}

Despite the considerable phenomenological success of inflation, much remains unknown about the microscopic physics that gave rise to it.  The observed (near) scale-invariance of the power spectrum points to a fairly symmetric ultraviolet theory: over a large number of e-folds, the Hubble parameter and other quantities must have been very nearly constant.  However, many ultraviolet constructions of inflation contain ingredients that can cause much larger violations of these symmetries.  For example, there can be additional particles, strings or branes whose couplings to the inflaton lead to unacceptably large violations of scale invariance.  By arranging the pieces appropriately, we are able to find viable models of inflation (see e.g.~\cite{Lyth:1998xn, Yamaguchi:2011kg, Baumann:2014nda} for review). Nevertheless, one might hope that these arrangements do not need to be terribly delicate and that even a haphazard (random) distribution of these ingredients would be capable of producing a viable model.    

In contrast, the qualitative appearance of many real-world systems is a poor guide to the short distance symmetries.  Many materials appear to be homogeneous on large scales despite the presence of microscopic inhomogeneities from impurities or irregular configurations of atoms.  These types of (statistical) deviations from perfect homogeneity are very well studied in the context of {\it disorder} (see e.g.~\cite{Lee:1985zzc,Belitz:1994zz} for review) and have revealed some remarkable properties, including the absence of propagating waves (i.e.~Anderson localization~\cite{PhysRev.109.1492, PhysRevLett.42.673}).  

An analogy between disorder in materials and the complexity of microscopic models of the universe has been advocated by several authors \cite{Tye:2007ja,Podolsky:2008du, Denef:2011ee,Anninos:2011kh}.  This work is often motivated by the complexity of vacua that would be necessary to explain the small size of the cosmological constant.  In the specific application to inflation, the analogue of disorder is to replace a nearly uniform potential with one that is generated randomly~\cite{Tegmark:2004qd, Aazami:2005jf, Easther:2005zr, Tye:2008ef,Frazer:2011tg,Agarwal:2011wm, Battefeld:2012qx,McAllister:2012am, Marsh:2013qca}.  This approach has proven valuable in answering statistical questions about inflation in specific random landscapes.  However, we are still laking an understanding of randomness in inflation akin to the case of solids.  Furthermore, inflation may not be of the slow-roll type and we should explore the space of models more generally.

More broadly, we would like to learn about the microphysics of inflation directly from cosmological data.    The most common and successful strategy to date has been targeted searches, where one constructs templates for well-motivated models and fits them to data.  This is typically the optimal way to look for a specific model, but one also hopes that other models will have significant overlap with these templates (perhaps in the sense of~\cite{Babich:2004gb}).  However, when the microphysics is itself random, there may be no expected signal in any one search.  Yet, if we examine all the searches in totality, one may expect to see statistically significant deviations from $\Lambda$CDM.  Exploring these types of models may lead to new ways to looking at the data for signatures of new physics.  

In this paper, we will explore these issues by further developing the connection between disorder and inflation.  When phrased in terms of the effective field theory (EFT) of inflation \cite{Creminelli:2006xe,Cheung:2007st}, there is very little difference between disorder in inflation and disorder in a real world material\footnote{Inflation and solids can be expressed in terms of spontaneously broken translations in time or space respectively.  Disorder in both cases is then defined as random coupling functions in these EFTs.  The difference between the two cases is whether these are functions of time (inflation) or space (solids).}.  We will use this similarity for guidance in deriving the formalism and it will lead us quite far from previous work on random potentials.  Our interest is both in identifying universal features that result from random microphysics and testing these ideas with current and future data.  

Using this formalism, we will find that disorder introduces additional noise into observable correlation functions.  This noise exhibits correlations that can be predicted for a given model; yet, there is no specific signal that is expected in correlation functions currently constrained by Planck~\cite{Ade:2013ydc}.  This noise arises from two separate physical effects: a random modulation of the amplitude of the metric fluctuations and the excitation of their quantum state.  Both of these effects could have been anticipate in terms of resonant features in each realization of the potential~\cite{Chen:2006xjb, Chen:2008wn, Pahud:2008ae, Flauger:2009ab,  Flauger:2010ja} and non-slow-roll generalization thereof~\cite{Behbahani:2011it,Behbahani:2012be}.

From a purely phenomenological perspective, disorder introduces a statistical field that modulates the power spectrum and/or bispectrum without additional violations of isotropy.  As such, it differs qualitatively from other types of modulations that are constrained by Planck~\cite{Ade:2013nlj,Ade:2013ydc} (or measured, in the case of the lensing potential~\cite{Smith:2007rg,Das:2011ak,vanEngelen:2012va,Ade:2013tyw}).  Nevertheless, the formalism of Hanson and Lewis~\cite{Hanson:2009gu} applies directly to this statistical field and it is straightforward to find the quadratic maximum likelihood estimator.  We discuss some of its properties for cosmic variance limited modes.  Interestingly, we find that constraints can be derived from general consistency tests of the data such as the $\chi^2$ of the fit to $\Lambda$CDM or the total integrated bispectrum~\cite{cmbbispectrum}.  Significant improvements in these constraints should be possible with a dedicated treatment.

This paper is organized as follows: in Section~\ref{sec:formalism} we will introduce the formalism of disorder to inflation and discuss the relation to previous work.  In Section~\ref{sec:noisy}, we will compute the effects of disorder on the power spectrum and bispectrum and their covariance matrices.  We perform these calculations in the context of several specific models of single-field inflation, but we expect the qualitative behavior to be more general.  In Section~\ref{sec:obs}, we discuss the observational consequences from a more phenomenological perspective, illustrating how constraints on the microscopic parameters can be derived.  We conclude in Section~\ref{sec:disc}.

This paper contains five appendices.  Appendix~\ref{app:slow} contains the details of the calculations relevant to disorder in single-field slow-roll inflation.  In Appendix~\ref{app:mono}, we explore disorder in the frequency domain.  This relation is important for understanding a number of the results in Section~\ref{sec:noisy}.  In Appendix~\ref{app:multi}, we illustrate how the formalism can be applied to multi-field inflation.  In Appendix~\ref{app:anderson}, we explain the relationship between our results and some aspects of Anderson localization.  In Appendix~\ref{app:estimator}, we derive the estimator for the disorder using the results of~\cite{Hanson:2009gu}.

\section{Formalism}\label{sec:formalism}

The idea of generating inflationary potentials at random is not new (see e.g.~\cite{Tegmark:2004qd, Aazami:2005jf, Easther:2005zr, Tye:2008ef,Frazer:2011tg,Agarwal:2011wm, Battefeld:2012qx, McAllister:2012am, Marsh:2013qca}).  Many studies have defined a distribution from which $V(\phi)$ is drawn and then determine the subsequent evolution for various initial conditions.  In these distributions, the variation of $V(\phi)$ is often large, even to the point where inflation may or may not occur from realization to realization.  These studies are well suited to address the likelihood that inflation can occur in a given random landscape and the statistical distributions of observables that result.

Here we will consider a mild introduction of randomness, analogous to introducing a potential 
\beq
V(\phi) = \bar V(\phi) + \delta V(\phi) \ ,
\eeq
where $\bar V$ is a fixed background potential and $\delta V$ is chosen at random.  Inflation and its basic observables are controlled by $\bar V$ and do not change with each realization.  We will consider more general variations on this idea, but the spirit of our construction will follow from here.

The motivation for this construction is analogous to disorder in solids.  In that case, one considers a material that is more or less uniform but, due to impurities (for example), there are small variations from point to point.  Since the locations of the impurities cannot be predicted, one treats them as a realization of a random distribution.

In essence, we want to allow for the possibility that the inflationary background is influenced by ``impurities" that appear randomly along a the path of the inflaton\footnote{Our results will not assume that inflation is described by slow-roll or even a fundamental scalar, but the language will occasionally be useful for explanatory purposes.}.  In addition, we will further assume that these impurities do not have light degrees of freedom associated with them and only modify the evolution of the background.  This assumption is typically categorized as {\it quenched disorder}.

We will analyze this problem using the EFT of inflation \cite{Cheung:2007st} as it offers several advantages.  First, there is a significant computational advantage as we do not have to determine the evolution of $\phi(t)$ for each realization of $V(\phi)$.  Second, this approach naturally allows for generalizations that are not described by slow-roll inflation.  Finally, this language is most similar to disorder in solids which will allow us to make the comparison, when applicable.

\subsection{The EFT of Inflation}

The EFT of inflation~\cite{Creminelli:2006xe,Cheung:2007st} describes the spontaneous breaking of time translations (which are then gauged by coupling to gravity).  The effective Lagrangian may contain explicit functions of $t$, provided they appear in the combination $t+\pi(\x,t)$ where $\pi$ is the Goldstone boson that non-linearly realizes the time translation symmetry.  To describe the inflationary background, the theory is coupled to gravity and the time translation symmetry becomes a component of the diffeomorphsims.  The gauge invariant action for the coupled system is given by
\bea\label{equ:EFT}
S &=& \int d^4 x \sqrt{-g} \Big[ \tfrac{1}{2} \Mp^2 {\cal R} + \Mp^2 \dot H(t+\pi) \partial_\mu (t + \pi) \partial^\mu (t+\pi) -\Mp^2 (3H^2(t+\pi)+\dot H(t+\pi))  \nonumber \\
&& + \tfrac{1}{2} M_2^4(t+\pi) [\partial_\mu (t + \pi) \partial^\mu (t+\pi)+1]^2 + \tfrac{1}{3!}M^4_3(t+\pi)  [\partial_\mu (t + \pi) \partial^\mu (t+\pi)+1]^3\, \Big] \ ,
\eea
where we have dropped terms that are higher order in $\partial_\mu(t+\pi)$ or in derivatives.  Solving Einsteins equations produces an FRW solution where $H(t) = \frac{\dot a}{a}$ with the parameters $H(t)$ and $M_{2,3}(t)$ being arbitrary\footnote{The null energy condition demands that $\dot H<0$ and the absence of superluminal modes requires $M_2^4> 0$.  These constraints are not explicitly built into the EFT of inflation but can be imposed as additional constraints. } functions of time. Inflation, as we will define it, is the case where the background is nearly de Sitter, namely $ | \dot H | \ll H^2$.

We can also identify the terms in this action around a slow-roll background using Einstein's equations : $\Mp^2 \dot H = -\tfrac{1}{2} \dot \phi(t)^2$ and $\Mp^2(3  H^2 + \dot H )= V(\phi(t))$.  Therefore, to make contact with inflation on a random potential, we should draw $H(t)$ from a probability distribution\footnote{The relation to disorder in a solid can be understood as follows.  A solid spontaneously breaks translations in space \cite{Leutwyler:1996er,Son:2005ak,Dubovsky:2005xd} and disorder is described by disorder potentials, which are stochastic functions of position but are independent of time.  Inflation spontaneously breaks time translations and the ``disorder potentials", $H(t)$ and $M_{2,3}(t)$, are stochastic functions of time but are independent of position.}.  We will also consider generalizations of this idea where $M_{2,3} (t)$ are stochastic variables.

The action for $\pi$ simplifies in the decoupling limit, $\Mp \to \infty$ and $\dot H \to 0$ holding $\Mp^2 |\dot H| \gg H^4$ fixed.  In this limit, the coupling the gravity becomes negligible and we can write the action directly for $\pi$ around an FRW background (up to total derivatives) as
\beq\label{equ:decoupling}
S = \int d^4x a^3 [ \Mp^2 ( \dot H(t) + \ddot H \pi) \partial_\mu \pi \partial^\mu \pi + 2 M_2^4(t)(\dot \pi^2 - \dot \pi \partial_\mu \pi \partial^\mu \pi) + 2\dot M_2^4(t) \pi \dot \pi^2  -\tfrac{4}{3} M_3^4(t) \dot\pi^3 ] \ ,
\eeq
where we have dropped terms ${\cal O}(\pi^4)$ and those suppressed by $\dot H \to 0$.  Typically, one will rewrite $M_2^4$ in terms of the speed of sound as $M_2^4(t) = - \frac{\Mp^2 \dot H (1-c_s^2)}{2 c_s^2}$.  For our purposes, it will be useful to work in terms of $M_2^4(t)$ directly.

For single-field inflation, observational predictions are computed in terms of the conserved curvature perturbation $\zeta$.  The action in (\ref{equ:decoupling}) and $\zeta = - H \pi$ will be sufficient for computing the correlation functions of $\zeta$ of interest in the next section.  Of course, we will have to justify the use of the decoupling limit in any such calculation, which requires showing that corrections of order $\dot H$ are negligible.

\subsection{Disorder}

We are now ready to introduce disorder into inflation.  Starting from (\ref{equ:decoupling}), we will split
\beq
\Mp^2 \dot H(t) \to \Mp^2 \dot H(t) + \Mp^2 \dot h(t) \qquad M_{2,3}^4(t) \to M_{2,3}^4(t) + m_{2,3}^4(t) \ ,
\eeq
where $\dot H(t), M_{2,3}^4(t)$ are fixed functions of time while $\dot h(t), m_{2,3}(t)$ are stochastic variables that will be sampled from some probability distribution.  The contributions from the stochastic variables are assumed to be sub-dominant to the fixed background and therefore it is reasonable to consider them as perturbations.

We now split the Hamiltonian into a solvable piece plus a perturbation, $\H = \H_0+\H_{\rm I}$, where $\H_{\rm I}$ is treated perturbatively and includes all of our stochastic variables and non-linear terms.  We want to compute the expectation value of some operator $Q(\tau)$ which is a product of local operators at different points in space but at a fixed conformal time $\tau$.  The in-in formalism tells us that for a given realization of the stochastic parameters, the quantum mechanical expectation value is given by \cite{Weinberg:2005vy}
\beq\label{equ:inin}
\langle Q(\tau) \rangle = \frac{\langle 0  | \left[ \bar T \exp(i \int^\tau_{-\infty (1+i \epsilon)} a d\tau' \H_{\rm I}(\tau') \right] Q_{\rm I}(\tau) \left[ T \exp(-i \int^\tau_{-\infty (1-i \epsilon)} d\tau' \H_{\rm I}(\tau') \right] | 0 \rangle }{\langle 0  | \left[ \bar T \exp(i \int^\tau_{-\infty (1+i \epsilon)}a d\tau' \H_{\rm I}(\tau')\right] \left[ T \exp(-i \int^\tau_{-\infty (1-i \epsilon)} a d\tau' \H_{\rm I}(\tau') \right] | 0 \rangle} \ ,
\eeq 
where $Q_I(\tau)$ is the interaction picture operator which is evolved with $\H_0$.  The denominator in this formula is likely unfamiliar for good reason.  First, when $\epsilon \to 0$, the denominator is simply $\langle U^{\dagger} U \rangle  = 1$ where $U$ is the unitary time evolution operator.  Even for finite $\epsilon$, one can check that the denominator is independent of $\tau$ and is therefore just an overall normalization which is irrelevant in most circumstances.  However, this constant will depend on the stochastic parameters which is why it can play a non-trivial role for disorder.

Ultimately, we want to compute the statistical predictions over many realizations of the stochastic variables:
\beq\label{eqn:ravg}
\langle Q(\tau) \rangle _R \equiv \left( \prod_i \int {\cal D} x_i(t) \right) P [x_i(t)] \langle Q(\tau) \rangle \ ,
\eeq
where $x_i(t)$ denotes the stochastic parameters $\dot h(t)$ and $m_{2,3}^4(t)$, which are sampled from a probability distribution $P[x_i(t) ]$.  The denominator in (\ref{equ:inin}) adds a new challenge to defining this theory non-perturbatively, as we compute the average over the stochastic parameters in~(\ref{eqn:ravg}) after taking the ratio in~(\ref{equ:inin}).  This is an important distinction that separates disorder from dissipation.

In practice, we will find that the denominator is negligible when performing perturbative calculations. We will expand out the numerator and denominator in powers $x_i$ and evaluate at each order in this expansion using the statistical correlation functions of $x_i(t)$ and quantum mechanical correlations for the fields, which factorize.  In this approach, the denominator removes some quantum mechanical vacuum bubble diagrams.  At low loop order, these diagrams are effectively trivial and can be removed by hand.  

Without loss of generality, we can make the assumption that
\beq
\langle  x_i(t) \rangle_R  = 0 \ ,
\eeq
as we can always shift the values of $\dot H(t)$ or $M_{2,3}^4(t)$ to absorb any non-zero average.  This ensures that the linear order correction to any correlation function will vanish.  Therefore, the leading contribution in $x_i(t)$ with $\H_{\rm I} = a^3(t) \sum_i x_i(t) {\cal O}_i(t)$ is given by
\bea
\langle Q(t) \rangle^{(2)}_R &=&  \sum_{i,j} \Big( \int^\tau_{-\infty} a^4(\tau_1) d \tau_1 \int^{\tau}_{-\infty} a^4(\tau_2) d\tau_2 \langle \cO_i(\tau_1)  Q(\tau)  \cO_j(\tau_2) \rangle  \langle x_i(\tau_1) x_j(\tau_2) \rangle_R \label{equ:quad} \\
&& - 2 {\rm Re}   \int^\tau_{-\infty} a^4(\tau_1) d \tau_1 \int^{\tau_1}_{-\infty} a^4(\tau_2) d\tau_2 \langle \cO_i(\tau_1) \cO_j(\tau_2) Q(\tau)   \rangle   \langle x_i(\tau_1) x_j(\tau_2) \rangle_R  \Big)\ . \nonumber
\eea
where $\tau \sim -\frac{1}{a H}$ is the conformal time.  If we specify all the correlations of the $x_i(t)$, then this procedure can be carried out to any order.  It is common in the literature to assume that the statistics of the disorder potentials are gaussian, in part because it can be treated non-perturbatively.  

The formalism for computing $\langle Q(t) \rangle_R$ is essentially the same as the one used for dissipation in the EFT of Inflation~\cite{LopezNacir:2011kk,LopezNacir:2012rm} (the basic formalism itself has a much longer history including~\cite{Feynman:1963fq, PhysRevLett.46.211, GellMann:1992kh, Boyanovsky:1994me, Berera:1996nv, Goldberger:2005cd, Porto:2007qi, Meade:2008wd} and was applied to inflation in~\cite{Berera:1996nv,Berera:1998px}).  In that case, one couples $\pi$ to some additional operators $\tilde{\cal O}(t, \x)$ with specified correlation functions.  Dissipation differs in detail because $\tilde{\cal O}(t,\x)$ has quantum mechanical fluctuations that depend on both time and space.  The first important difference this introduces is that disorder does not contain terms linear in $\pi$, such as ${\cal L} \supset \tilde x_1(t) \nabla^2(t+\pi)$, as they correspond to tadpoles\footnote{Alternatively, we could include such terms in the action but they will only modify the $\k=0$ mode of $\pi$.  Such terms can always be removed by a diffeomorphism.} for $\pi$ around any realization of the stochastic parameters.  Linear terms are allowed for dissipation (e.g.~$\tilde x_1(t) \to \tilde{\cal O}(t,\x)$) because the quantum fluctuations of $ \tilde{\cal O}$ eliminate the tadpole, provided that $\langle \tilde \cO \rangle =0$.  The second distinction is that for quenched disorder there is no feedback (response) between $\pi$ and $x_i(t)$.  This can be understood simply from symmetries: since $x_i(t)$ is only a function of time it, cannot depend on $\pi(\x,t)$ locally.  

So far, we have been completely agnostic about the nature of the disorder parameters.  The assumption we will make here is that their distributions are independent and purely local, such that 
\beq\label{equ:xixj}
\langle x_i (t) x_j (t') \rangle_R \propto \delta_{ij} \delta(t-t') \to \langle x_i (\tau) x_j (\tau') \rangle_R = C_{i} 
\delta_{ij} (-H \tau)^{p+1} \delta(\tau-\tau') \ .
\eeq
Here we have introduce a power law in conformal time, $(-H \tau)^p$, to allow for some breaking of scale invariance.  In the limit $p \to 0$, this two point function is invariant under $t \to t+c$ and our results will be scale invariant.  In addition, to simplify calculations it is useful to analytically continue in $p$, even when taking $p \to 0$ at the end of the calculation.  There is no reason that $p$ need be the same for each $i$, but in what follows this generalization can be implemented trivially.  

It should come as no surprise that the stochastic variables $\dot h(t)$ and $m_{2,3}^4(t)$ multiply derivatives of $\pi$.  Since $\pi$ is a Goldstone boson, the underlying symmetry puts strong constraints on the action.  Non-derivative terms are constrained by tadpole cancelation and do not arise in the decoupling limit.  It is therefore natural to expect that the stochastic terms are irrelevant in the technical sense.  When we consider time scales much shorter than a Hubble-time (i.e. the modes are inside the horizon), we can see this more precisely by using (\ref{equ:xixj}) since $\langle x_i^2 \rangle_R$ scales as $t^{-1}$.  This suggests that we can treat $x_i(t)$  an ``operator" of dimension $1/2$.  In the same limit, $\pi$ behaves as a dimension one field and therefore we have that $\dot h(t) \partial_\mu \pi \partial^\mu \pi$, $m_2^4(t) \dot \pi^2$ and $m_3^4(t) \dot \pi^3$ are dimension $9/2$, $9/2$ and $13/2$ respectively.  In this sense, disorder is irrelevant during inflation.

When the modes cross the horizon, this scaling behavior breaks down and results in a correction of fixed amplitude.  The size of the effect is given as ratio of scales between $H$ and some scale $\Lambda_i \gg H$ which controls the strength of the irrelevant operator.  If we were to take $H \to 0$ holding everything else fixed, disorder should have no effect on inflationary observables.  For example, if
\beq
\langle m_2^4(\tau )  m_2^4(\tau') \rangle_R = \frac{1}{\Lambda_2} (\Mp^2 |\dot H|)^2 (-H \tau) \delta(\tau-\tau') 
\eeq
then we will find that at ${\cal O}(m_2^8)$, disorder produces corrections of order $\frac{H}{\Lambda_2}$ as one would expect from a dimension $9/2$ operator.  

It may seem surprising that disorder is irrelevant.  Specifically, in a slow-roll model we can write down a stochastic mass term for the inflaton which is relevant.  Such terms are present in the full $\pi$ Lagrangian in equation~(\ref{equ:EFT}) but they vanish in the decoupling limit $\dot H, \dot h \to 0$ (e.g.~${\cal L} \supset 6 \Mp^2 |\dot H| \dot h \pi^2$).  Taking this limit with fixed $H^2$ amounts to assuming that the background solution produces a large number of e-folds of inflation.  Essentially, demanding that inflation occurred at all requires that any relevant disorder parameters are negligible\footnote{If we were to work with $\zeta$ rather than $\pi$, one would find that all the relevant disorder parameters are absent~\cite{Maldacena:2002vr}.} and that only the irrelevant terms survive.  

While our discussion is focused on disorder in single field inflation, the formalism naturally generalizes to multi-field inflation as shown in Appendix~\ref{app:multi}.  Additional fields are not as constrained by the underlying symmetry and relevant disorder does arise.  Nevertheless, even relevant disorder in inflation does not exhibit the more dramatic consequences of disorder seen in solids, as discussed in Appendix~\ref{app:anderson}.  As long as disorder is perturbative, we expect that most of our qualitative conclusions will generalize beyond the single-field case.  

Given the apparent generality of these arguments, it is worth emphasizing that there were two key assumptions that separate the present work from the work of many previous authors.  First, we are working in the limit of perturbative disorder.  Inherent to the above power counting is the assumption that the dominant contribution to inflationary correlation functions is from a non-stochastic component.  Many previous studies assumed the full inflationary background was generated stochastically and therefore one cannot rely on perturbative power counting techniques to estimate the size of the effect.  Our power counting shows that the effects of disorder can be made controllably small but has little to say if we made the effect large from the beginning.

The second difference is that we have assumed the correlation length of the stochastic field is essentially zero\footnote{The model studied in \cite{Tye:2008ef} shares some qualitative features with ours, including short-range correlations for a stochastic (multi-field) potential.}.  If we draw the coefficients of the lagrangian from a distribution, it is likely that there are long range correlations in the potential (i.e.~stochastic parameters separated by a Hubble-time are correlated).  There was nothing about our formalism that demanded we make this choice, but short range disorder is the most analogous to the case of solids which was one of our primary motivations.  Allowing for long range correlations is an interesting generalization of the results presented here.

\section{Noisy Correlation Functions}\label{sec:noisy}

Having introduced disorder, we are now ready to compute corrections to various correlation functions.  Around a given realization, the power spectrum and bispectrum will be modified at leading order in the stochastic parameters.  However, these effects average to zero over many realizations.  Nevertheless, the implication is that there will be added noise in these correlation functions or, alternatively, there are additional contributions to their covariance matrices.  As a result, it will be natural to consider both the correlation function and its covariance matrix at the same time.  

Throughout this section, we will assume that the stochastic parameters obey
\beq\label{equ:stoch2pt}
\langle x_i(\tau) x_j(\tau') \rangle_R =\delta_{ij} \frac{\Mp^4 |\dot H|^2}{\Lambda_i} (-H \tau)^{p+1} \delta(\tau-\tau') \ ,
\eeq
where $x_i =\{ \Mp^2 \dot h, m_2^4\}$.  The assumption that the different parameters are uncorrelated means that the leading corrections can be considered in isolation.  We also typically assume that the statistics are gaussian, but we will briefly consider the non-gaussian case as well.

Under the above assumption, the corrections from stochastic terms in (\ref{equ:decoupling}) relative to leading non-stochastic terms are suppressed by $\frac{H}{\Lambda_i} \ll 1$.  The stochastic terms that we have neglected by taking the decoupling limit are further suppressed by at least an additional factor of $\frac{|\dot H|}{H^2} \ll 1$ and are therefore negligible.  Non-stochastic slow-roll corrections may be comparable to the leading stochastic term, i.e. $\frac{H}{\Lambda_i} \sim \frac{|\dot H|}{H^2}$, but can be treated independently at the order we are working.

\subsection{Noisy Power Spectra}\label{sec:power}

The most basic observable of interest in cosmology is the power spectrum.  One can think of the effects of $\dot h$ and $m_2^4$ as a stochastic modification of the amplitude of the fluctuations, so it should be no surprise that these contribute extra noise in the power spectrum (i.e.~a non-gaussian trispectrum).  In addition to the modulation of the amplitude, the time-dependence of any realization will also excite the quantum state of $\pi$.  Both effects should be familiar from the context of resonance~\cite{Chen:2006xjb, Chen:2008wn, Pahud:2008ae, Flauger:2009ab,  Flauger:2010ja, Behbahani:2011it,Behbahani:2012be}, which is closely related to disorder, as we show in Appendix~\ref{app:mono}.
\vskip 8pt
\noindent {\bf Power spectrum from $m_2^4$:}\hskip 6pt Let us begin by computing corrections from $2 m_2^4 \dot \pi^2$ with the further simplification\footnote{Under these assumptions, a Lorentz invariant UV completion (i.e.~a UV theory with vanishing commutators outside the light-cone) cannot literally produce this EFT as $c_s > 1$ on realizations of $m_2^4$.  This can be avoided with a non-zero $M_2^4$.  Since the fluctuations of $m_2^4$ are small compared to $\Mp^2 \dot H$, the required modification of $c_s$ is likely negligible. Nevertheless, this example is primality for illustration and we will not worry about this detail.} that $M_2^4 = 0$.   With this assumption, we have $\pi_{\k} = \bar \pi_{\k} \hat a_{\k}^{\dagger}+ {\rm h.c.}$ where $\hat a_{\k}^{\dagger}$ is the creation operator and 
\beq
\bar \pi_{\k}= \frac{H}{2 \Mp |\dot H|^{1/2} } \frac{1}{k^{3/2}} (1-i k \tau) e^{i k \tau} \ .
\eeq
This case is relatively simple because $\H_{\rm I}$ only involves
\beq
\dot{\bar \pi}_{\k}= - \frac{H^2}{2 \Mp |\dot H|^{1/2}} \tau^2 k^{1/2} e^{i k \tau} \ .
\eeq
Plugging into (\ref{equ:quad}) we have that
\bea
 \Delta \langle \zeta_{\k} \zeta_{\k'} \rangle'_R &=& P_\zeta(k) 2 k^2(1+k^2 \tau_0^2) \int_{-\infty}^{\tau_0} d\tau\int_{-\infty}^{\tau_0} d\tau'e^{2 i k (\tau - \tau')} \frac{\langle m_2^4(\tau) m_2^4(\tau')\rangle_R}{\Mp^4 \dot H^2} \nonumber \\
&&- 2\, {\rm Re}\, P_\zeta(k) 2 k^2 (1+i k\tau_0)^2 e^{-2i k \tau_0}  \int_{-\infty}^{\tau_0} d\tau\int_{-\infty}^{\tau} d\tau'  e^{2 i k \tau'} \frac{\langle m_2^4(\tau) m_2^4(\tau')\rangle_R}{\Mp^4 \dot H^2} \ ,\label{equ:m2powb}
\eea
where $\langle \rangle'$ means we have removed the momentum conserving delta function.  In writing this expression to have kept the $i\epsilon$ prescription implicit (see Appendix~\ref{app:slow} for details).  Using (\ref{equ:stoch2pt}) and taking the limit $\tau_0 \to 0$, one finds that the first line vanishes for $p > -2$.  Integrating the second line we get
\bea
\Delta \langle \zeta_{\k} \zeta_{\k'} \rangle'_R &=& - \, {\rm Re}\, P_\zeta(k) 2 k^2 (1+i k\tau_0)^2 e^{-2i k \tau_0}  \int_{-\infty}^{\tau_0} d\tau \frac{(-H \tau)^{p+1}}{\Lambda_2}e^{2 i k \tau} \\
&\to&P_\zeta(k) \frac{H}{\Lambda_2} \frac{\cos(\tfrac{p \pi}{2} )  \Gamma[p+2]}{2^{1+p}} \left( \frac{H}{k}\right)^p \equiv \Delta P_\zeta(k) \ ,
\eea
where we took $\tau_0 \to 0$ in the second line.  As expected, the correction is suppressed by $\tfrac{H}{\Lambda_2}$ and is scale invariant in the limit $p \to 0$.  The only subtle aspect of this calculation is that the $\delta$-function in (\ref{equ:stoch2pt}) is evaluated at the boundary of integration in (\ref{equ:m2powb}), which effectively introduces a factor of $\tfrac{1}{2}$.

\vskip 8pt
\noindent {\bf Trispectrum from $m_2^4$:}\hskip 6pt Around a specific  $m_2^4(t)$, the power spectrum is modified at linear order by
\bea
\Delta \langle \zeta_{\k} \zeta_{\k'} \rangle' &=& P_\zeta(k) 2{\rm Re} (-i) \int^{\tau_0}_{-\infty}  d \tau k e^{2 i k \tau} \frac{m_2^4(\tau)}{\Mp^2 |\dot H|} \ .
\eea
Of course, this term averages to zero, but it means that the amplitude of the power spectrum varies randomly as a function of $k$.  We would expect this to show up as more noise in the measurement of the power spectrum or equivalently, as added power in the 4-point function.  This is straightforward to compute as
\bea
\langle \zeta_{\k_1} \zeta_{\k_2} \zeta_{\k_3} \zeta_{\k_4} \rangle'_R &=& (P_\zeta + \Delta P_\zeta)(k_1)   (P_\zeta + \Delta P_\zeta)(k_3) \delta(\k_1 + \k_2) \delta(\k_3 +\k_4) \times  \label{eqn:power1}\\
&&\left[1-{\rm Re}\frac{k_1 k_3 }{4} [  \int^{\tau_0}_{-\infty} d\tau \frac{(-H \tau)^{p+1}}{\Lambda_2} ( e^{2 i (k_1+k_3) \tau} -  e^{2 i (k_1-k_3) \tau}) \right]+{\rm permutations}  \nonumber \\
&\to& (P_\zeta + \Delta P_\zeta)(k_1)   (P_\zeta + \Delta P_\zeta)(k_3) \delta(\k_1 + \k_2) \delta(\k_3 +\k_4) \times  \label{equ:tri1}\\
&&\left[1+ \frac{H}{\Lambda_2} \frac{\cos(\tfrac{p \pi}{2} )  \Gamma[p+2]}{2^{4+p}} \left(\frac{H^p k_1k_3}{(k_1+k_3)^{2+p}} - \frac{H^p k_1k_3}{|k_1-k_3|^{2+p}} \right) \ \right]+{\rm permutations}  \ ,\nonumber
\eea
where we simplified the expression from the beginning by eliminating contributions that vanish as $\tau_0 \to 0$. This expression is only valid to leading order in $\frac{H}{\Lambda_2}$.  Notice that the first line alone is the gaussian expectation given the power spectrum $P_\zeta + \Delta P_\zeta$.

At first sight, the appearance of two $\delta$-functions may suggest that this contribution is not ``connected".  Specifically, this shows that there is no exchange of momentum between the two pairs of fields.  Nevertheless, the trispectrum is connected (i.e.~it is irreducible) due to the exchange of energy.  Specifically, the pairs are correlated through the coupling to $m_2(\tau)$ which depends explicitly on time but not on space.  This type of behavior is perfectly consistent because this is a non-relativistic system.

There is something very clearly wrong with (\ref{equ:tri1}) in the limit $k_1 \to k_3$.  We see that the second term diverges (which, a priori, is not necessarily an issue) and that is negative.  However, $k_1 = k_3$ should be computing the diagonal elements of the covariance matrix, which one would expect to be positive on very general grounds.  Something unphysical is happening in this limit.

The first term in (\ref{equ:tri1}) captures the intuitive effect that the amplitude of the power spectrum is varying randomly in time.  The second term is capturing the excitation from the ground state, which characteristically introduces divergences at $k_1 =k_3$.  The energy at which the state can be excited is related to the timescale on which the background varies.  By using $\langle m_2^4(\tau) m_2^4(\tau') \rangle \propto \delta(\tau-\tau')$, we have implicitly allowed for arbitrarily rapid changes in the stochastic parameters which means arbitrarily large energies in the state.  This is shown most directly in Appendix~\ref{app:mono} in terms of a resonance model.  To maintain control, we impose a bound on the frequencies that appear in the stochastic parameters, $x_i(t) \supset e^{i \omega t}$, such that $\omega < \bar \Lambda < (\Mp^2 |\dot H|)^{1/4}$~\cite{Behbahani:2011it,Flauger:2013hra}.  Fourier transforming this to the time domain introduces an additional suppression factor $e^{\bar \epsilon (k_1+k_3) \tau}$ in equation (\ref{eqn:power1}), where $\bar \epsilon \equiv H / \bar \Lambda$ (see Appendix~\ref{app:mono} for details).  Talking the limit $p \to 0$, one finds
\bea
\langle \zeta_{\k_1} \zeta_{\k_2} \zeta_{\k_3} \zeta_{\k_4} \rangle'_R &\to& (P_\zeta + \Delta P_\zeta)(k_1)   (P_\zeta + \Delta P_\zeta)(k_3) \delta(\k_1 + \k_2) \delta(\k_3 +\k_4) \times  \label{equ:tri2}\\
&&\left[1+ \frac{H}{64 \Lambda}  \left(\frac{ k_1k_3}{(k_1+k_3)^{2}} + {\rm Re} \frac{ k_1k_3}{(i (k_1-k_3) +  \bar \epsilon (k_1+k_3) )^{2}} \right) \ \right]+{\rm permutations} \ .\nonumber
\eea
This modified result now has a more physical limit $k_1 \to k_3$, as it is bounded and positive.  The above formula will be modified depending on how the short-time behavior is resolved.  In practice, the resolution is important only for $k_1\sim k_3$, at which point the result is essentially determined by scale invariance up to the overall normalization (in the limit $p \to 0$).  

\vskip 8pt
\noindent {\bf Power spectrum in Slow-Roll:}\hskip 6pt  Computing corrections to the power spectrum during slow-roll inflation (i.e.~contributions from $\Mp^2 \dot h(t)$) are somewhat more complicated than the above case and have been computed in Appendix~\ref{app:slow}.  The calculation itself is similar in structure to the previous case, with the final result being
\beq
\Delta  \langle  \zeta_{\k} \zeta_{\k'}  \rangle'_R = - \frac{H}{\Lambda_1} P_{\zeta} (k) \, \frac{4+(1-p)p}{2-p}   \cos(\tfrac{p \pi}{2} )\Gamma[p] \Big(\frac{H}{2 k }\Big)^p \ . \label{eqn:pneq0}
\eeq
Unlike the case of $m_2^4$, the contribution for $\dot h$ is not well behaved in the limit $p \to 0$,
\beq
\lim_{\p \to 0} \Delta \langle  \zeta_{\k} \zeta_{\k'}  \rangle'_R =- 2\frac{H}{\Lambda_1} P_{\zeta} (k) \, [\frac{1}{p} + \frac{3}{4} - (\gamma + \log \frac{k}{2 H} ) ] \ .
\eeq
We see that the result diverges as $p \to 0$ and the finite part is not scale invariant.  The formula as written in equation~(\ref{eqn:pneq0}) holds for $p>0$ but is not valid in the $p \to 0$ limit.  It is instructive to understand the source of the problem.

First, it is clear that any violation of scale invariance with $p = 0$ must be accompanied by a divergence.  We started with an integral that was manifestly invariant under the rescaling $k \to \lambda k$ and $\tau \to \lambda^{-1} \tau$.  The conservation of $\zeta$ further guarantees that our result should be independent of $\tau_0$~\cite{Weinberg:2003sw}.  Therefore the only way that the result could violate scale invariance (i.e~contains explicit functions of $k$) is if the integral itself is not well defined.

The cause of the divergence is again due to exited states of arbitrarily high energy. The time integral implicitly sums over all the modes that are in excited states, which is unbounded.  When this divergence is a power law, it is easily removed through analytic continuation in $p$.  However, for $p = 0$ the divergence is logarithmic and cannot be removed trivially. We can again regulate the integral by including a suppression factor $e^{ \bar \epsilon k \tau}$ and, taking the $p \to 0$ limit, one finds that
\bea
\lim_{p\to 0} \Delta \langle  \zeta_{\k} \zeta_{\k'}  \rangle'_R = \frac{1}{2}  \frac{H}{\Lambda_1} P_\zeta(k) [-3 - 2 \log \frac{ \bar \epsilon^2}{4}] \ .
\eea
The result is now scale invariant, as it should be.  We also see that the result is divergent as $\bar \epsilon \to 0$, which is consistent with the appearance the unphysical behavior we observed in the absence of the regulator.

\vskip 8pt
\noindent {\bf Trispectrum in Slow-Roll:}\hskip 6pt Unlike the power spectrum, the subtleties associated with excited states and the trispectrum were already visible with the simpler case studied above.  The expression for general $p$ is given the appendix~\ref{app:slowtri}.  The result is rather lengthly and not terribly illuminating.  Talking the limit $p \to 0$ (and dropping the gaussian piece) produces the result 
\bea
\Delta \langle \zeta_{\k_1} \zeta_{\k_2} \zeta_{\k_3} \zeta_{\k_4} \rangle'_R &\to&  \frac{H}{64 \Lambda_1} P_\zeta(k_1)   P_\zeta (k_3) \delta(\k_1 + \k_2) \delta(\k_3 +\k_4) \times  \label{equ:tri2}\\
&&  \left(\frac{-16 k_1^2 + 31 k_1 k_3 - 16 k_3^2 }{ (k_1-k_3)^2}+ \frac{16(k_1^2 +k_3^2)}{k_1 k_3} {\rm ArcTanh}(\tfrac{k_3}{k_1} ) \right)  +{\rm permutations} \nonumber 
\eea
for $k_1 \geq k_3$.  As we saw before, this expression is badly behaved when $k_1 \to k_3$.  The divergence that appears in this limit are cut off when $k_1 - k_3 < (k_1+k_3) \bar \epsilon$, as we saw in (\ref{equ:tri1}).  Unlike the power spectrum, the trispectrum is scale invariant without the appearance of $\log \bar \epsilon$ corrections.

\subsection{Noisy Bispectra}

As we saw in the previous section, the primary signature of disorder is that it introduces additional noise into the power spectrum.  This effect is most easily captured as a modification to the trispectrum beyond the usual gaussian expectation.  There is no reason for such effects to be limited to the power spectrum.  As an illustrative example, we will show two ways in which disorder can produce noisy bispectra.
\vskip 8pt
\noindent {\bf Disordered Interactions:}\hskip 6pt  It was clear from the EFT description that there is no reason that disordered couples only to $\pi$ quadratically.  The most straightforward example comes from
\beq
{\cal L}_{\rm int} = \tfrac{1}{3!}m^4_3(t+\pi)  [\partial_\mu (t + \pi) \partial^\mu (t+\pi)+1]^3\,  \simeq - \tfrac{4}{3}m^4_3(t) \dot \pi^3  \ , \label{equ:m3def}
\eeq
where $m_3^4(t)$ is a gaussian random field with a power spectrum 
\beq\label{equ:m3}
\langle m_3^4(\tau) m_3^4(\tau') \rangle_R = \frac{\Mp^6 |\dot H|^3}{H^4 \Lambda_3} (-H \tau)^{p+1} \delta(\tau-\tau') \ .
\eeq
For any realization of $m_3^4(t)$, this interaction introduces a bispectrum.  As with the power spectrum, at linear order in $m_3^4(t)$, the average amplitude of the bispectrum is zero.  Unlike the power spectrum, there is no contribution to the bispectrum at order $(m_3^4)^2$.  Nevertheless, at this order, we find a non-zero correction to the covariance of the bispectrum (beyond the gaussian contribution) which, in the $p\to0$ limit, is given by
\bea
\Delta \langle \zeta_{\k_1} \zeta_{\k_2} \zeta_{\k_3} \zeta_{\k_4}  \zeta_{\k_5} \zeta_{\k_6} \rangle_R &=& \frac{10}{3} \frac{H}{\Lambda_3} P_\zeta(k_1) P_{\zeta}(k_2) P_{\zeta}(k_3) \delta(\k_1+\k_2+\k_3) \delta(\k_4+\k_5+\k_6) \times  \\
&&\frac{k_1^2 k_2^2 k_3^2}{k_3 k_4 k_5} \Big( \frac{1}{|\sum_i k_i|^6 } - \frac{1}{|k_1+k_2+k_3 - k_4-k_5-k_6|^6 }  \Big)+{\rm permutations} \nonumber \ .
\eea
The form of the bispectrum covariance is very similar to the trispectrum we found in (\ref{equ:tri1}).  The second term shows a divergence when $k_1 +k_2 +k_3 = k_4 + k_5 + k_6$ (or any permutation thereof).  At this point, it should be clear that this is the result of being in an excited state of arbitrarily large energy.  This divergence is regulated if we cutoff in energy being input into the system.

If we consider the limit when $k_1 +k_2 +k_3 \to k_4 + k_5 + k_6$, there is still important information on the particular configurations where the noise dominates.  We regulate the divergence at energies above $\bar \Lambda$ by including a suppression factor $e^{\tfrac{1}{2} \bar \epsilon (\sum_i k_i) \tau}$.  To leading order one finds
\beq
\lim_{k_4+k_5+k_6 \to k_1 +k_2+k_3} \Delta \langle \zeta_{\k_1} \zeta_{\k_2} \zeta_{\k_3} \zeta_{\k_4}  \zeta_{\k_5} \zeta_{\k_6} \rangle_R' =\frac{10}{3} \frac{H}{\bar \epsilon^6 \Lambda_3}  \frac{(2 \pi^2 \Delta_\zeta^2)^3}{k_1 k_2 k_3 k_4 k_5 k_6 |k_1+k_2 +k_3|^6} \ ,
\eeq
where $\Delta_\zeta^2 = 2.2 \times 10^{-9}$~\cite{Ade:2013zuv}.  We should compare this to the shape of the bispectrum that is generated by ${\cal L} \supset \frac{4}{3}M_3^4 \dot \pi^3$~\cite{Senatore:2009gt}
\beq
B_{\dot \pi^3} (k_1 , k_2,k_3) = \frac{486}{5}\frac{1}{k_1 k_2 k_3 |k_1 + k_2 +k_3|^3} \ .
\eeq
Therefore, we see that
\beq
\lim_{k_4+k_5+k_6 \to k_1 +k_2+k_3} \Delta \langle \zeta_{\k_1} \zeta_{\k_2} \zeta_{\k_3} \zeta_{\k_4}  \zeta_{\k_5} \zeta_{\k_6} \rangle_R \propto B_{\dot \pi^3} (k_1 , k_2,k_3) B_{\dot \pi^3} (k_4, k_5,k_6)  \ .
\eeq
It should not be surprising that we have excess noise in the shape $B_{\dot \pi^3} (k_1 , k_2,k_3)$, given that this was the form of the interaction.  This observation is still non-trivial as the time dependence of the coefficients modify the shape, which did give rise to the unusual contributions to the covariance matrix away from the limit $k_1 +k_2 +k_3 \to k_4 + k_5 + k_6$.
\vskip 8pt
Disordered interactions can also produce a contribution to the trispectrum by essentially contracting two of the external momenta in the 6-point function to form an internal line.  Because this exchanges momentum, the form of the trispectrum has a more familiar $\delta$-function structure  
\bea
\Delta \langle \zeta_{\k_1} \zeta_{\k_2} \zeta_{\k_3} \zeta_{\k_4}  \rangle_R &=& 30 \frac{H}{\Lambda_3} P_\zeta(k_1) P_{\zeta}(k_2) \delta(\k_1+\k_2+\k_3+\k_4)  \times \label{equ:trifrombi} \\
&&\frac{k_1^2 k_2^2 |\k_1 +\k_2|}{k_3 k_4} \Big( \frac{1}{|\sum_i k_i|^6 } - \frac{1}{|k_1+k_2-k_3 - k_4|^6}  \Big)+{\rm permutations} \nonumber \ .
\eea
The first term in this expression is insensitive to $\bar \epsilon$ and has a cosine~\cite{Regan:2010cn} of $0.25$ with the constant trispectrum, defined by
\beq
\langle \zeta_{\k_1} \zeta_{\k_2} \zeta_{\k_3} \zeta_{\k_4}  \rangle =  t_{\rm NL} \frac{8 (2 \pi^2 \Delta_\zeta^2)^3}{(k_1 k_2 k_3 k_4)^{9/4} }  \ .
\eeq
The current limit is given by $t_{\rm NL} = (-1.33 \pm 3.62) \times 10^6$~\cite{Fergusson:2010gn}.  A proper analysis of this trispectrum is beyond the scope of this work, but we should expect to derive a constraint $\frac{H}{\Lambda_3} \lesssim 10^{-2}$ or better.  
\vskip 8pt
\noindent {\bf Non-gaussian Disorder:}\hskip 6pt In the presence of disorder, non-gaussianity in $\pi$ may arise from non-gaussianity in the disorder field.  As an illustrative example, we will consider the case where
\beq
\langle m_2^4(\tau_1)m_2^4(\tau_2) m_2^4(\tau_3) \rangle_R = \frac{(\Mp^2 |\dot H|)^3}{\Lambda_{4}^2} (-H \tau_1)^{p+2} \delta(\tau_1 - \tau_2) \delta (\tau_2 - \tau_3) \label{equ:ngnoise}\ .
\eeq
We again compute the bispectrum covariance because the bispectrum itself vanishes at every order in $m_2^4$.  We now must compute to order $(m_2^4)^3$.  To do so, we use
\bea
\langle Q(t) \rangle^{(3)}_R &=&   16 {\rm Re}\Big( i \int^\tau_{-\infty} a^4(\tau_1) d \tau_1 \int^{\tau_1}_{-\infty} a^4(\tau_2) d\tau_2\int^\tau_{-\infty} a^4(\tau_3) d \tau_3 \nonumber \\ && \qquad \qquad \times \langle \cO_2(\tau_1)\cO_2(\tau_2)   Q(\tau)  \cO_2(\tau_3) \rangle  \langle m_2^4(\tau_1) m_2^4(\tau_2) m_2^4(\tau_3) \rangle_R \label{equ:cube} \\
&& -   i  \int^\tau_{-\infty} a^4(\tau_1) d \tau_1 \int^{\tau_1}_{-\infty} a^4(\tau_2) d\tau_2 \int^{\tau_2}_{-\infty} a^4(\tau_3) d\tau_3  \nonumber \\ && \qquad \qquad \times\langle \cO_2(\tau_1) \cO_2(\tau_2) \cO_2(\tau_3)Q(\tau)   \rangle   \langle m_2^4(\tau_1) m_2^4(\tau_2) m_2^4(\tau_3) \rangle_R  \Big)\ . \nonumber
\eea
The calculation is again straightforward (being careful with delta functions at the boundaries of integration) and results in
\bea
\Delta \langle \zeta_{\k_1} \zeta_{\k_2} \zeta_{\k_3} \zeta_{\k_4}  \zeta_{\k_5} \zeta_{\k_6} \rangle_R &=& -\frac{1}{32} \frac{H^2}{\Lambda_4^2} P_\zeta(k_1) P_{\zeta}(k_3) P_{\zeta}(k_5) \delta(\k_1+\k_2) \delta(\k_3+\k_4)\delta(\k_5+\k_6) \times \nonumber \\
&& \Big(\frac{k_1 k_3 k_5}{|k_1+k_3 - k_5|^3 } +\frac{1}{3} \frac{k_1 k_3 k_5}{|k_1+k_3 +k_5|^3 }  \Big)+{\rm permutations}  \label{equ:ngnoise2}\ .
\eea
The unusual appearance of $\delta$-functions is again due to the ``exchange" of stochastic fields which do not exchange momentum.

\section{Observational Signatures}\label{sec:obs}

The calculations presented in the previous section show how disorder affects the statistics of $\zeta_{\k}$.  Unlike more traditional signatures, the dominant features of these models are similar in form to the covariance matrices for the cosmic variance limit modes.  As a result, these signatures may hide more easily in data than for more traditional observables.  

In this section, we will discuss the predictions of these models from a phenomenological perspective.  The general implications of this framework are not overly sensitive to the specific models presented in the previous section.  For the power spectrum, this will be closely related to existing work on statistical anisotropy of the cosmic microwave background (CMB), as demonstrated in Appendix~\ref{app:estimator}.  The signal for the bispectrum is qualitatively similar to the power spectrum, but is less directly related to the signatures of existing models.

Unfortunately, we will not derive precise constraints on our model parameters, not even suboptimal ones.  We will explain the order of magnitude of constraints that can be derived, in principle, using existing analyses.  However, both the power spectrum and bispectrum do show some additional noise, which is the signal of these models.  However, these are likely due to instruments effects and approximations in the covariance matrices, rather a primordial signal\footnote{We thank Raphael Flauger for explaining these points.}.  We will estimate the constraints assuming that there is no such additional noise (for cosmic variance limited modes) but deriving a precise limit would require a careful treatment of these issues.

\subsection{Noisy Power Spectra}

In order to gain intuition for the observational signatures, it is useful to think about the modulation of a mode, $\zeta_\k$, by a given realization of $m_2^4(t)$.  Using the in-in formalism, or the equations of motion, we can write $\zeta_\k = \bar\zeta_\k \hat a_\k^{\dagger}+ {\rm h.c.}$ with
\beq
 \bar \zeta_{\k} = \bar \zeta_{\k, 0} [1 + f(k) ] \ ,
\eeq
where $ \bar \zeta_{\k, 0}$ is the solution computed with $m_2^4 = 0$ and 
\beq\label{equ:fk}
f(k) =  \int^0_{-\infty} d\tau \, k \sin(k \tau) \frac{2 m_2^4(\tau)}{ \Mp^2 |\dot H|} e^{ i k \tau} \ .
\eeq
We were able write a simple expression for (\ref{equ:fk}) because there is no issue taking $\tau_0 \to 0$ before integration (this would not hold for $\dot h(t)$).  Any realization of $m_2^4(t)$ will then impact the temperature fluctuations using
\beq
\Theta_{\ell, m} = 4\pi i^\ell \int \frac{d^3 k}{(2\pi)^3} \Delta_\ell(k) \Big(\bar \zeta_{\k, 0} [1 + f(k) ]\hat a_\k+ {\rm h.c.} \Big) Y_{\ell, m}(\hat k) \ ,
\eeq
where $\Theta = \frac{\delta T}{T}$, $\Delta_\ell(k)$ is the linear CMB transfer function and $Y_{\ell,m}$ are the spherical harmonics.  We see that $f(k)$ will play the role of an additional statistical field that modulates the temperature fluctuations.

The problem of how to reconstruct an arbitrary (but small) modulating field from observations of the CMB temperature multipoles was solved by Hansen and Lewis \cite{Hanson:2009gu}.  The connection is more clear in terms of the covariance
\beq\label{equ:kappa}
C^{\theta \theta }_{\ell,m; \ell',m'} = C_\ell  \delta_{\ell \ell'} \delta_{m m'} (1+ \kappa_\ell) + C^{N N}_{\ell,m; \ell',m'}
\eeq
where $C^{N N}_{\ell,m; \ell',m'}$ is the contribution from instrumental noise and
\beq
\kappa_\ell \equiv\frac{4\pi}{C_\ell} \int d \log k \, \tfrac{ k^3}{2 \pi^2} P_\zeta(k) \Delta_\ell(k)^2 \, 2{\rm Re} f(k) \ .
\eeq
The form of the modulating field $\kappa_\ell$ is unlike the lensing potential or additional fields during inflation, which introduce off-diagonal terms in $m$ and $m'$.  The reason no such effects arise is because every realization of $m_2^4(\tau)$ preserves homogeneity and isotropy.  As a result, the temperature power spectrum is diagonal in $\ell$ and $m$ for any $m_2^4(\tau)$.  Now we can use the results of Hanson and Lewis to write an estimator for $\kappa_\ell$ (see Appendix~\ref{app:estimator} for a derivation).  The form of the estimator most intuitive when we consider only cosmic variance limited modes (i.e.~${\bf C}^{N N}=0$), in which case we find
 \beq\label{equ:estimator}
\hat \kappa_\ell =\frac{1}{C_\ell}  \Big (C_\ell^{\rm obs.}-C_{\ell} \Big)  \ ,
\eeq
where $\frac{1}{2 \ell +1}\sum_m \Theta^*_{\ell,m} \Theta_{\ell,m} \equiv C_\ell^{\rm obs.}$.  The form of the estimator is essentially what one would have defined starting from equation~(\ref{equ:kappa}).  It is just difference between the observed $C_\ell$ and the one predicted when $\kappa_\ell =0$.

A proper analysis will be left to future work, but we can estimate the size of constraint that could be derived from existing analyses.  Suppose we take our reconstructed $\kappa_\ell$ and compute
\beq
\sum_\ell \tfrac{1}{2} (2 \ell+1)\hat\kappa_\ell^2 = \sum_\ell \frac{(2 \ell +1) }{2 } \frac{(C_\ell^{\rm obs.} - C_\ell)^2}{C_\ell^2} =  \chi^2 \ .
\eeq 
We get a constraint on a sum over $\kappa_\ell^2$ from the $\chi^2$ of the fit to the $\Lambda$CDM model.  Of course, there is noise in our estimator so we have
\beq
\langle \hat \kappa_\ell \hat \kappa_\ell \rangle_R = \frac{2}{2 \ell +1} [1 +  \langle \kappa_\ell^2 \rangle_R] 
\eeq
which implies that
\beq
 \langle \chi^2 \rangle_R = \ell +\sum_{\ell} \tfrac{1}{2}(2 \ell+1) \langle \kappa_\ell^2 \rangle_R \ .
\eeq
The appearance of the factor of $(2\ell +1)$ in accounts for the increasing precision we expect in measurements of $C_\ell$ (or $\kappa_\ell$) at higher values of $\ell$.  In other words, additional noise in the higher $\ell$ modes are more highly constrained because there are more modes.  

There is some subtlety to measuring the value of $\chi^2$ correctly, but it is consistent with $\Lambda$CDM at the $\chi^2 - \ell ={\cal O}(10-100)$ level~\cite{Ade:2013kta}.  From here is it straightforward to derive constraints on the underlying parameters.  For example, if $ \langle \kappa_\ell^2 \rangle_R = \kappa_0^2$, then we would find
\beq
\kappa_0 \lesssim 10^{-2} \ ,
\eeq
using Planck cosmic variance limited modes up to $\ell_{\rm max} = 1500$.  This constraint is somewhat weaker than the naive expectation of $N_{\rm modes}^{-1/2} \simeq 10^{-3}$ (where $N_{\rm modes}\simeq \ell_{\rm max}^2$ is the number of cosmic variance limited modes observed) because of we allowed for some excess $\chi^2$ that in the existing data.  

Using $\chi^2$ to derive limits on disorder will typically lead to a weaker constraint than expected.  The main difficulty is that it requires a very precise model for the expected noise, as any excess $\chi^2$ will weaken the constraint (or fake a signal). On the other hand, the noise produced by $\kappa_\ell$ should also exhibit correlations between different values of $\ell$ that are predicted from the trispectra computed in Section~\ref{sec:power}.  These correlations are not accounted for by $\chi^2$ but could be used to separate disorder for other sources of excess noise.  A given fit may find a $\chi^2$ that differs from gaussian predictions for any number of reasons.  However if we were to observe the specific correlations in the noise predicted by disorder, it might be more difficult to find a conventional explanation.  

\subsection{Noisy Bispectra}

Technically speaking, there is very little distinction between the signature in the power spectrum and in the bispectrum.  We could repeat the previous steps to construct an estimator for the field that modulates the bispectrum.  For the case of non-gaussian disorder, this estimator is still given by equation~(\ref{equ:estimator}) but we should also look for higher order correlations in the noise.  For disordered interactions, we would likely need a new estimator constructed from the bispectrum directly.  In principle, such analyses can be performed, but are quite different from what has been currently constrained.

In practice, our best understanding of primordial bispectra comes from projecting the data onto specific bispectrum templates.  These templates may be organized into a basis of orthogonal polynomials which then provide a complete basis for the space of signals \cite{Fergusson:2009nv}.  The coefficient of each template is measured in the data and can but used to reconstruct the bispectrum of the CMB \cite{cmbbispectrum}.

To understand what our noisy bispectra would look like to such a procedure, we will imagine that observations are made directly in terms of $\zeta_{\k}$ (the application to the CMB is straightforward, but beyond the scope of this work).  To measure the amplitude in a given shape, $\bar B(k_1,k_2, k_3)$, we use the estimator
\beq
{\hat f}_{\rm NL} = \frac{1}{\bar B \cdot \bar B} \sum_{\k_i} \zeta_{\k_1} \zeta_{\k_2} \zeta_{\k_3} \bar B(k_1,k_2,k_3) [P_\zeta(k_1) P_{\zeta}(k_2) P_{\zeta}(k_3)]^{-1} \ ,
\eeq
where 
\beq
\bar B \cdot \bar B \equiv  \sum_{\k_i}  \bar B(k_1,k_2,k_3)^2 [P_\zeta(k_1) P_{\zeta}(k_2) P_{\zeta}(k_3)]^{-1} \ .
\eeq
If we average over realizations of the stochastic parameters, we will find that $\langle {\hat f}_{\rm NL} \rangle_R = 0$.  The noise that is expected is given by
\beq
\langle {\hat f}_{\rm NL}^2 \rangle_R = \frac{6}{\bar B \cdot \bar B} +\frac{1}{(\bar B \cdot \bar B)^2} \sum_{\k_i , \k_j'} \frac{ B(k_1,k_2,k_3) B(k_1',k_2',k_3') \Delta \langle \zeta_{\k_1} \zeta_{\k_2} \zeta_{\k_3} \zeta_{\k_1'} \zeta_{\k_2'} \zeta_{\k_3'} \rangle_{R}}{P_\zeta(k_1) P_{\zeta}(k_2) P_{\zeta}(k_3)P_\zeta(k_1') P_{\zeta}(k_2') P_{\zeta}(k_3')} \label{equ:fnlnoise} \ .
\eeq
The first term is what you would expect from a purely gaussian universe, while the second term depends only on the stochastic contribution.

At this point, one could plug in specific models and templates into equation~(\ref{equ:fnlnoise}) to determine the numeric coefficients as a function of the stochastic variables.  What should be clear is that the expected noise in $\hat f_{\rm NL}$ will be larger than would arise in a purely gaussian universe.

However, if all we did was measure a single $\hat f_{\rm NL}$, then we would only derive a relatively weak bound on this additional noise (which may or may not be a meaningful bound on the microscopic parameters).  If the correction to the noise is small, there would be slightly higher probability to measure large values of $f_{\rm NL}$.  However, this is not useful if you only make one measurement.  In order to strengthen the constraint, we want to measure $\hat f_{\rm NL}$ for a number of different templates.  In essence, this gives us access to many more realizations.

For example, suppose we measure the amplitude of $n$ orthogonal templates labelled by $\hat f_{\rm NL, i}$, and compute
\beq\label{equ:fnlsum}
\Delta \bar f_{\rm NL}^2 \equiv \frac{1}{n} \sum_{i=1}^n \hat f_{\rm NL,i}^2  =  \langle \hat f_{\rm NL}\rangle_R^2 + {\cal O}(\frac{1}{\sqrt{n}} ) \ ,
\eeq
where we have assumed that each template is normalized to have the same error, $\langle \hat f_{\rm NL}^2\rangle_R$. The second term is the error in $\Delta \bar f_{\rm NL}^2$ itself, which vanishes in the limit $n \to \infty$ as $\frac{1}{\sqrt{n}}$.  This allows us to make a meaningful measurement of the additional noise in the bispectrum and therefore constrain stochastic contributions to the bispectrum.

Planck essentially performs the above measurement in the CMB in terms of the modal decomposition.  Specifically, they construct a basis of orthogonal templates and measure the amplitude of each template.  They show results for $n = 600$ templates and they define a similar quantity to (\ref{equ:fnlsum}) which they call the total integrated bispectrum \cite{cmbbispectrum}.  They find some excess beyond the gaussian expectation but it is not statistically significant (see figure 10 of \cite{Ade:2013ydc}).  From the above scaling, this will constrain the additional noise to be below the 5 percent level.

Although the above is intuitively clear from gaussian statistics, it may not be completely transparent how this translates into a bound on a specific model.  The first term in (\ref{equ:fnlnoise}) scales as $\Delta_\zeta^{-2} N_{\rm modes}^{-1}$ (where $N_{\rm modes}$ are the number of data points used to measure each individual template) but the scaling of the second term depends in detail on the template and the noise.  In the presence of a non-zero bispectrum, the second term would scale as $N_{\rm modes}^0$, which we can interpret as the usual $(S/N) \propto N_{\rm modes}^{-1/2}$.  For disordered interactions, this scaling is weakened but depends on the correlation length in the noise (which is ultimately model dependent).  We also saw that a constraint is derivable from the trispectrum in equation~(\ref{equ:trifrombi}).  To determine which constraint is stronger is a detailed question.  In practice, it may be easier to construct an estimator for the noise in the bispectrum, much as we did for the power spectrum.  We will leave a detailed analysis for later work.

In the case of non-gaussian disorder, the $N_{\rm mode}$ dependence of both terms is similar (due to the additional $\delta$-functions).  However, unless the disorder is highly non-gaussian, we expect the stronger constraint to arise from the power spectrum.

\section{Discussion}\label{sec:disc}

In this paper, we introduced a formalism for disorder to inflation.  In the presence of disorder, the evolution of the background is subject to perturbations that arise randomly.   This problem is treated by introducing stochastic functions of time into the effective theory of inflation.  We computed observational predictions in the limit where the disorder can be treated perturbatively and is uncorrelated at different times.

The most robust observational prediction is the presence of additional noise in the power spectrum or bispectrum.  This noise is correlated between different scales, which can be predicted for a microscopic model.  These correlations should allow one to distinguish disorder from other sources of noise or errors in the covariance matrices.  In this paper, we showed how constraints could be derived from existing analyses, but a dedicated analysis is an interesting problem for future work.

The results presented here differ from other studies of inflation on a random potential in a few key respects.  Previous work has typically focused on multi-field models of inflation where the couplings constants are drawn from a random distribution.  In most cases, the resulting potentials cannot be treated as a small perturbation around a fixed model and are therefore in the regime of strong disorder.  In other cases, the sizes of these random coefficients were chosen to be small and may admit a perturbative treatment.  However, it is likely that when these models are phrased in terms of disorder, their stochastic fields will have a large correlation length.  This may explain why previous models have not shown a tendency for noisy correlation functions (although see~\cite{Tye:2008ef} for a possible exception).

In the absence of a microscopic model, the choices we made were motivated by the similarity to the treatment of disorder in solids.  These choices may have a more natural origin in specific models such as trapped inflation~\cite{Green:2009ds, D'Amico:2012ji} or solid inflation~\cite{Gruzinov:2004ty, Endlich:2012pz, Sitwell:2013kza}.  In trapped inflation, the microscopic origin of the model involves a number of independent particle production events at fixed positions along the inflaton trajectory.  The locations were chosen to be evenly spaced to avoid significant violations of scale invariance.  As we have seen here, moderate statistical variations in the location of these events would likely produce signatures similar to what was found above (although we would also need to include dissipation~\cite{LopezNacir:2011kk,LopezNacir:2012rm}).

Solid inflation is perhaps the most natural home for disorder.  Because inflation is literally driven by a solid, the results from previous work on disorder should apply directly.  One would simply be postulating that inflation is driven by an amorphous solid or a solid with impurities.  The observational signatures in that case will differ qualitatively from those we found here, because any realization of the disorder (which would be a function of position, rather than time) would break homogeneity and isotropy.

Disorder is a very general framework that should have applications well beyond the narrow choices we made for simplicity or lack of imagination.  There are likely interesting generalization in the single and multi-field contexts that will show different behavior  from what we illustrated here and perhaps also novel signatures for current or future probes of the initial conditions.  

\subsubsection*{Acknowledgements}

We thank Daniel Baumann, Richard Easther, Raphael Flauger, Jonathan Frazer, Sean Hartnoll, Justin Khoury, Liam McAllister, Daan Meerburg, Michael Mulligan, Alberto Nicolis, Riccardo Penco, Rafael Porto and Alex van Engelen  for helpful discussions.  This work was supported by a NSERC Discovery Grant.

\newpage
\appendix

\section{Corrections in Slow-Roll}\label{app:slow}

The leading stochastic term in slow-roll inflation is $\Mp^2 \dot h(t) \, \partial_\mu \pi \partial^\mu \pi$.  While this may be the correction of most interest, the calculation itself is more involved.  In this appendix we will provide a detailed derivation of the results in the main text.

The primary challenge with slow-roll is that we must split $\partial_\mu \pi \partial^\mu \pi = - \dot \pi^2 + \tfrac{1}{a^{2}} \partial_i \pi \partial^i \pi$ because time and space are treated differently.  In practice, it is easiest to treat each term a separate perturbation, but it also means there are more terms to compute.  

The corrections any equal time correlator at order $\dot h^2$ can be computed using (\ref{equ:quad}) if we defined ${\cal O}_1 \equiv  (\dot \pi^2) $, ${\cal O}_2 \equiv - \tfrac{1}{a^{2}} \partial_i \pi \partial^i \pi$ and $x_1(t) = x_2(t) \equiv \Mp^2 \dot h(t)$.  We will therefore break the corrections into four terms
\bea\label{equ:qij}
\langle Q(t) \rangle_{R,\alpha \beta} &=&  \int^\tau_{-\infty} a^4(\tau_1) d \tau_1 \int^{\tau}_{-\infty} a^4(\tau_2) d\tau_2 \langle \cO_\alpha(\tau_1)  Q(\tau)  \cO_\beta(\tau_2) \rangle  \langle x_\alpha(\tau_1) x_\beta(\tau_2) \rangle_R \\
&& - 2 {\rm Re} \Big(  \int^\tau_{-\infty} a^4(\tau_1) d \tau_1 \int^{\tau_1}_{-\infty} a^4(\tau_2) d\tau_2 \langle \cO_\alpha(\tau_1) \cO_\beta(\tau_2) Q(\tau)   \rangle   \langle x_\alpha(\tau_1) x_\beta(\tau_2) \rangle_R  \Big)\ , \nonumber
\eea
where $\alpha, \beta \in \{1,2\}$.  
\subsection{Power Spectrum}
We have already computed $\langle \zeta_{\k} \zeta_{\k'} \rangle'_{R,11}$ in Section~\ref{sec:power}, where we found
\beq
\Delta \langle \zeta_{\k} \zeta_{\k'} \ \rangle_{R,11} =\frac{1}{2} \frac{H}{\Lambda_1}  P_\zeta(k) \frac{\cos(\tfrac{p \pi}{2} )  \Gamma[p+2]}{4} \left( \frac{H}{2 k}\right)^p
\eeq  
The next term to compute is
\bea
\Delta\langle  \zeta_{\k} \zeta_{\k'} \rangle'_{R,12}+ \Delta \langle  \zeta_{\k} \zeta_{\k'}  \rangle'_{R,21} &=&  2 {\rm Re}\Big[ P_\zeta(k) \frac{k^2(1+k^2 \tau_0^2)}{2} \int_{-\infty}^{\tau_0}   \frac{(-H \tau)^{p-1}}{\Lambda_1} (1-i k \tau)^2 \Big]\\
&&- \, {\rm Re}\Big[ \, P_\zeta(k) \frac{k^2 (1+i k\tau_0)^2 e^{-2i k \tau_0}}{2} \\ &&\times  \int_{-\infty}^{\tau_0} d\tau  \frac{(-H \tau)^{p-1}}{\Lambda_1} e^{ i 2 k \tau} \left( (1-i k \tau)^2 +(1+k^2 \tau^2)  \right) \Big] \nonumber
\eea
Evaluating the terms separately in the limit $\tau_0 \to 0$ (assuming $p>-1$) we get
\bea
&& 2 {\rm Re}\Big[ P_\zeta(k) \frac{k^2(1+k^2 \tau_0^2)}{2} \int_{-\infty}^{\tau_0}   \frac{(-H \tau)^{p-1}}{\Lambda_1} (1-i k \tau)^2 \Big] \to \frac{H}{\Lambda_1} P_\zeta(k) \times \frac{-(-H \tau_0)^p}{p} 
\eea
and
\bea
&&- \, {\rm Re}\Big[ \, P_\zeta(k) \frac{k^2 (1+i k\tau_0)^2 e^{-2i k \tau_0}}{2}  \times  \int_{-\infty}^{\tau_0} d\tau  \frac{(-H \tau)^{p-1}}{\Lambda_1} e^{ i 2 k \tau} \left( (1-i k \tau)^2 +(1+k^2 \tau^2)  \right) \Big] \nonumber \\
&& \qquad \qquad \to \frac{H}{\Lambda_1} P_\zeta(k) \times \left[ \frac{ (-H \tau_0)^p}{p}  - \frac{1}{2} (2+p)\cos(\tfrac{p \pi}{2} )\Gamma[p] \Big(\frac{H}{2 k }\Big)^p \right] \ .
\eea
Combining the two terms leads to
\bea
\Delta\langle  \zeta_{\k} \zeta_{\k'}  \rangle'_{R,12}+\Delta \langle  \zeta_{\k} \zeta_{\k'}  \rangle'_{R,21} &=& - \frac{1}{2}  \frac{H}{\Lambda_1} P_\zeta(k)   (2+p)\cos(\tfrac{p \pi}{2} )\Gamma[p] \Big(\frac{H}{2 k }\Big)^p \ .
\eea
Notice that when $p < 0$ the two contributions diverge as $\tau_0 \to 0$ but these divergences cancel.  This type of cancelation occurs in nearly all calculations and is necessarily to get sensible results\footnote{Every realization of $\dot h$ is a model of slow-roll inflation so $\zeta$ must be conserved outside the horizon~\cite{Maldacena:2002vr, Weinberg:2003sw}.  }.  However, this feature would not appear if we were to naively analytically continue in $\tau$ while keeping the correlations in $x_i(\tau)$ local.  Specifically, in the first term of equation~\ref{equ:qij}, the correct analytic continuation has ${\rm Im} \tau_1 >0$ and ${\rm Im } \tau_2 <0$ (this can be seen in equation~\ref{equ:inin}) and therefore a purely delta function correlation would vanish.  As a result, there would be no cancelation of the divergence in the second term and we would get an unphysical result.  Therefore, physical results in complex $\tau$ will require that $x_i(\tau)$ have non-local correlations.

Finally we need to compute $\langle Q(t) \rangle_{R,22}$.  This case is nearly identical to the previous case, but there are more divergent contributions that cancel between the two terms, leaving the final result
\bea
\Delta\langle  \zeta_{\k} \zeta_{\k'}  \rangle'_{R,22} \to- 2\frac{H}{\Lambda_1} P_{\zeta} (k) [\tfrac{1}{2-p} + p(5+p) ] \cos(\tfrac{p \pi}{2} )\Gamma[p] \Big(\frac{H}{2 k }\Big)^p \ .
\eea
Collecting all these terms we find that
\bea
\Delta\langle  \zeta_{\k} \zeta_{\k'}  \rangle'_R = -\frac{H}{\Lambda_1} P_{\zeta} (k) \, \frac{4+(1-p)p}{2-p}   \cos(\tfrac{p \pi}{2} )\Gamma[p] \Big(\frac{H}{2 k }\Big)^p \ .
\eea

In the limit $p \to 0$, we find the result is both divergent and violates scale invariance.  As discussed in Section~\ref{sec:power} and Appendix~\ref{app:mono}, these unphysical results arise from including changes to $\dot h(t)$ of arbitrarily large frequency.  Putting a cutoff in frequency introduces a suppression factor $e^{\bar \epsilon k \tau}$, which modifies the above results by terms proportional to $( \bar \epsilon k  )^p$.  For $p>0$, these terms are suppressed by $\bar \epsilon \ll 1$ and may be dropped.  However, in the $p \to 0$ limit, when we include these corrections we find
\bea
 \Delta\langle  \zeta_{\k} \zeta_{\k'}  \rangle'_{R,11} &=& \frac{1}{2}  \frac{H}{\Lambda_1} P_\zeta(k)  [\frac{1}{\bar\epsilon^2} + \frac{1}{4} ]\\
\Delta\langle  \zeta_{\k} \zeta_{\k'} \rangle'_{R,12}+ \Delta\langle  \zeta_{\k} \zeta_{\k'}  \rangle'_{R,21} &=& - \frac{1}{2}  \frac{H}{\Lambda_1} P_\zeta(k)  [\frac{2}{\bar \epsilon^2}  +1 + \log \frac{\bar \epsilon^2}{4} ] \\
\Delta\langle  \zeta_{\k} \zeta_{\k'} \rangle'_{R,22}  &=&  \frac{1}{2}  \frac{H}{\Lambda_1} P_\zeta(k) [\frac{1}{\bar \epsilon^2} - \frac{9}{4} - \log \frac{\bar \epsilon^2}{4}] \ .
\eea
Collecting all these terms we find that
\bea
\lim_{p\to 0} \Delta \langle  \zeta_{\k} \zeta_{\k'}  \rangle'_R = \frac{1}{2}  \frac{H}{\Lambda_1} P_\zeta(k) [-3 - 2 \log \frac{\bar \epsilon^2}{4}] \ .
\eea

\subsection{Trispectrum}\label{app:slowtri}

The trispectrum calculation follows nearly identical steps to the power spectrum, with only minor modifications.  We will therefore quote the results of the calculation with limited details.  The first term of interest is 
\bea
\Delta \langle \zeta_{\k_1} \zeta_{\k_2} \zeta_{\k_3} \zeta_{\k_4} \rangle_{R,11} &=& P_\zeta (k_1)   P_\zeta (k_3) \delta(\k_1 + \k_2) \delta(\k_3 +\k_4) \times  \label{equ:apptri1}\\
&&\left[\frac{H}{\Lambda_1} \frac{\cos(\tfrac{p \pi}{2} )  \Gamma[p+2]}{2^{6+p}} \left(\frac{H^p k_1k_3}{(k_1+k_3)^{2+p}} - \frac{H^p k_1k_3}{|k_1-k_3|^{2+p}} \right) \ \right]+{\rm permutations} \nonumber
\eea
As discussed in the main text, the divergence when $k_1 \to k_3$ is unphysical and requires some additional input.  We will quote results in this appendix with the understanding that $k_1 \neq k_3$.  From the discussion in section~\ref{sec:power}, it should be clear how to correct these formulas in the limit $k_1 \to k_3$.

The terms that are relevant for the slow-roll are
\bea
\Delta \langle \zeta_{\k_1} \zeta_{\k_2} \zeta_{\k_3} \zeta_{\k_4} \rangle_{R,12 + 21} &=&  \frac{H^{1+p}}{\Lambda_1} \frac{\cos(\tfrac{p \pi}{2} )  \Gamma[p]}{2^{5+p}} P_\zeta(k_1)   (P_\zeta)(k_3) \delta(\k_1 + \k_2) \delta(\k_3 +\k_4) \times \\
&&\Big[ \frac{2 (k_1-k_3)^2(k_1^2+k_3^2)- p k_1 k_3 (k_1-2 k_3)(k_3-2 k_1) +p^2 k_1^2 k_3^2 }{k_1 k_3|k_1-k_3|^{2+p}}  \nonumber \\
&& -\frac{2 (k_1+k_3)^2(k_1^2+k_3^2)+2 p k_1 k_3 (k_1+ k_3)^2 +p^2 k_1^2 k_3^2 }{k_1 k_3(k_1+k_3)^{2+p}}  \Big] +{\rm permutations} \nonumber
\eea
and
\bea
 \Delta\langle \zeta_{\k_1} \zeta_{\k_2} \zeta_{\k_3} \zeta_{\k_4} \rangle_{R,22 } &=&  \frac{H^{1+p}}{\Lambda_1} \frac{\cos(\tfrac{p \pi}{2} )  \Gamma[p]}{2^{6+p}} P_\zeta(k_1)   (P_\zeta)(k_3) \delta(\k_1 + \k_2) \delta(\k_3 +\k_4) \times \\
&&\Big[ \frac{4(2+p) (k_1^4+k_3^4) - 4 (2+p)^2 k_1 k_3 (k_1^2 +k_3^2)+ (1+p)(16 +p(6+p))k_1^2 k_3^2 }{(2-p)k_1 k_3|k_1-k_3|^{2+p}}  \nonumber \\
&& -\frac{4(2+p)(k_1+k_3)^2 (k_1^2+k_3^2+p k_1 k_3) + p(p+1)(2-p)k_1^2 k_3^2 }{(2-p)k_1 k_3|k_1+k_3|^{2+p}}  \Big] \nonumber \\&&+ \, \, {\rm permutations} \ . \nonumber
\eea
The sum of these terms gives the correction to the gaussian trispectrum, but no particular simplification occurs.  The results simplify significantly in the $p \to 0$ limit, where we find
\bea
\Delta \langle \zeta_{\k_1} \zeta_{\k_2} \zeta_{\k_3} \zeta_{\k_4} \rangle'_R &\to&  \frac{H}{64 \Lambda_1} P_\zeta(k_1)   P_\zeta (k_3) \delta(\k_1 + \k_2) \delta(\k_3 +\k_4) \times \\
&&  \left(\frac{-16 k_1^2 + 31 k_1 k_3 - 16 k_3^2 }{ (k_1-k_3)^2}+ \frac{16(k_1^2 +k_3^2)}{k_1 k_3} {\rm ArcTanh}(\tfrac{k_3}{k_1} ) \right)  +{\rm permutations} \nonumber 
\eea
assuming $k_1 > k_3$.

\section{Relation to Resonance} \label{app:mono}

The models of disorder we study here can be recast in terms of resonant behavior of the potential.  Specifically, there is a well studied class of models where a traditional slow-roll potential $V_0(\phi)$ is modified by~\cite{Chen:2006xjb, Chen:2008wn, Pahud:2008ae, Flauger:2009ab,Flauger:2010ja}
\beq
V(\phi) = V_0(\phi) + \Lambda^4 \cos(\frac{\phi}{f} ) \ .
\eeq
To leading order in slow-roll, the extra term modifies $ H^2(t) = H_0^2 [1 + b \cos(\omega t) ]$ for $\omega \sim \frac{\dot \phi}{f}$ and $b  \sim \Lambda^4/ (3 \Mp^2 H_0^2)$.  These sinusoidal terms have a natural generalization to any coupling in the EFT of inflation~\cite{Behbahani:2011it,Behbahani:2012be}.

We see that disorder and oscillatory features can be related if we make the choice
\beq
\Mp^2 \dot h = \int_{-\infty}^\infty \frac{d \omega}{2 \pi}  \lambda_\omega e^{i \omega t}
\eeq
were $\lambda^\dagger_\omega = \lambda_{-\omega}$ is a stochastic variable which satisfies 
\beq\label{eqn:freq}
\langle \lambda_{\omega} \lambda_{\omega'} \rangle_R = \frac{\Mp^4 \dot H^2}{\Lambda_1} (2 \pi) \delta(\omega+\omega') \ .
\eeq
After integrating over $\omega$, we recover the case $p= 0$ for $\dot h(t)$.  Clearly any model of disorder can be related to a model of resonance by taking the Fourier transform in this way.

The advantage of working with the oscillations is that it makes it more transparent that we are introducing arbitrarily large frequencies into the system.  In perturbation theory, these terms will excite modes of arbitrarily large energy.   This leads to a number of unphysical features of our results in the limit $p \to 0$.  Of course, there is only a finite energy density available so this is a sign that the effective description breaks down for these high frequencies.  In this language, it is clear that we can resolve this issue by enforcing introducing an exponential suppression in equation~(\ref{eqn:freq}) for $\omega >\bar \Lambda$.  With the added suppression, let us examine the previously problematic contribution to $\langle  \zeta_{\k} \zeta_{\k'}  \rangle'_{R,22}$ from
\bea
{\cal I} &\equiv& \int^\tau_{-\infty} a^4(\tau_1) d \tau_1 \int^{\tau}_{-\infty} a^4(\tau_2) d\tau_2 \langle \cO_2(\tau_1)   \zeta_{\k} \zeta_{\k'}(\tau_0) \cO_2(\tau_2) \rangle  \Mp^4 \langle \dot h(\tau_1) \dot h(\tau_2) \rangle_R \ .
\eea
We are interested in the regime where the delta function in not a good approximation, namely $-k \tau_{1,2} \gg \frac{\Lambda}{H}$.  Using $| \Delta \tau | \equiv |\tau_2 -\tau_1 | \ll | \tau_1 |$, we find that
\bea
{\cal I}  &\sim&\frac{1}{16}\frac{H}{\Lambda_1} P_{\zeta}  \int_{-\infty}^{\tau_0}  \frac{d \tau_1 }{(-H \tau_1)^4} (1+k^2 \tau_0^2)(1+k^2 \tau_1^2)^2 \int_{-\infty}^\infty d\, \Delta\tau \int_{-\infty}^{\infty} \frac{d \omega}{2 \pi}e^{-\frac{|\omega|}{ \bar \Lambda}} e^{i k \tau_{12}} e^{i \tfrac{\omega}{H} \frac{\Delta\tau}{\tau_1} } \nonumber \\
&\sim& \frac{1}{16}\frac{H}{\Lambda_1} P_{\zeta}  \int_{-\infty}^{\tau_0}  \frac{d \tau_1 }{(-H \tau_1)^4}  (1+k^2 \tau_0^2)(1+k^2 \tau_1^2)^2 (- H \tau_1) e^{ \bar \epsilon k \tau_1} \ .
\eea
where $\bar \epsilon \equiv H / \bar \Lambda$.  We see that the integral over $\tau_1$ is the same as the previous case when $\bar \Lambda \to \infty$, but is exponentially suppressed for $k |\tau_1| \gg \frac{\bar \Lambda}{H}$.  As shown in Section~\ref{sec:noisy}, including this exponential in $\tau$ is sufficient for achieving physical results in the $p\to 0$ limit.

In general, the form of this suppression will depend on the details of the model at small separation in time.  In the main text we used the above suppression factor, $e^{ \bar \epsilon k \tau_1}$, in part because it simplifies the calculations.  In practice, the observational predictions should depend weakly on this choice.  For higher point correlation functions, we will use $e^{\tfrac{1}{2}  \bar \epsilon  \tau_1 \sum_i k_i}$, where $k_i$ are the external momenta.

\section{Multi-field Generalization}\label{app:multi}
In the main text we consider disorder in single-field inflation.  The application to multi-field inflation is straightforward, but is complicated by the fact that it is less clear what quantities are observable.  In this appendix, we will illustrate how disorder can be introduced into a multi-filed context, but we will leave a detailed study to future work.

Multi-field inflation was first introduced into the EFT of inflation in \cite{Senatore:2010wk} with the application to quasi-single field inflation~\cite{Chen:2009zp} being future developed in \cite{Baumann:2011nk, Green:2013rd}.  The basic structure is the same as the single-field case but with additional fields that may couple to $\pi$.  The complication is that $\zeta$ may depend on all the fields, not just $\pi$.  As in the case of single field inflation, the role of disorder is to promote the coupling functions (of time) to stochastic variables.  For the purpose of illustration, we will include a scalar field $\chi$ that is light during inflation but is a spectator to the background dynamics (i.e. $\langle \chi \rangle = 0$).  For simplicity, we will be described the dynamics of $\chi$ by the action
\beq\label{eqn:chi}
S_\chi = \int dt d^3 x a^3 \left( -\tfrac{1}{2} \partial_\mu \chi \partial^\mu \chi - \tfrac{1}{2} m^2(t+\pi) \chi^2 \right) \ .
\eeq
We will take $m^2(t)$ to be a purely stochastic variable with
\beq
\langle m^2(\tau) m^2(\tau') \rangle_R = \bar m^3 (-H \tau) \delta(\tau-\tau') \ .
\eeq
Given the calculations that have been performed in terms of $\zeta$ and $\pi$, it should be clear how to compute correlation functions of $\chi$, using $\chi_\k = \bar\chi_\k a_\k^{\dagger} + {\rm h.c.}$ with
\beq
\bar \chi_\k = \frac{H}{\sqrt{2 k^3}} (1-i k \tau) e^{i k \tau} \ .
\eeq
The one new feature that arises in this case is that the corrections may depend on $\log \tau_0$.  This is the leading power of $\tau_0$ can arise in a perturbative equal-time correlation function, as shown in~\cite{Weinberg:2006ac}.  To illustrate this feature, it is sufficient to compute the correction to power spectrum
\beq
\Delta \langle \chi \chi \rangle_R = \frac{H^2}{2 k^3} \times \frac{\bar m^3}{H^3} \frac{14 - 4 \gamma - 4 \log (2 \tau_0 k)}{9} \ .
\eeq
In this case, the separate terms in equation~\ref{equ:quad} produce contributions that scale as $\tau_0^{-6}$, $\tau_0^{-4}$ and $\tau_0^{-2}$ but all such terms cancel in the final result.

In contrast to couplings in single-field inflation, this stochastic mass parameter is relevant.  Nevertheless, the calculation can be kept under perturbative control for sufficiently small $\bar m$ provided that $\frac{\bar m^3}{H^3} \ll 1$.  This concretely demonstrates that the is no analogy of Anderson localization for these additional fields (or at least that it doesn't occur for arbitrarily weak disorder).  A concrete comparison is made in Appendix~\ref{app:anderson}.

It is worth emphasizing that it is not a physical requirement that disorder is perturbative.  The specific coupling in (\ref{eqn:chi}) is precisely the one studied in \cite{Green:2009ds,LopezNacir:2011kk,LopezNacir:2012rm} and can lead to dissipation in the limit of strong disorder\footnote{We thank Rafael Porto for emphasizing this point.}.  This regime is also calculable, although not with the techniques used in the present work. Other regimes of strong disorder may show similarly interesting phenomenology but we currently lack the tools to explore them.


\section{Connection with Localization}\label{app:anderson}

Perhaps the most famous implication of disorder is Anderson localization~\cite{PhysRev.109.1492, PhysRevLett.42.673}.  Anderson localization is the phenomena that wave functions can become localized in materials as a result of disorder.  One interesting feature is that localization of electrons occurs even for arbitrarily weak disorder in $D \leq 2$ spatial dimensions.  We can roughly understand why the result depends on the spatial dimension as follows.  Suppose we have a fermion, $\psi$, coupled to a medium with impurities localized at random positions such that the action is given by
\beq
S_\psi = \int d^Dx dt [ i \bar \psi \slashed{\partial}\psi - m(\x) \bar \psi \psi ] \ ,
\eeq
where $m(\x)$ is a stochastic variable with $\langle m(\x) m(\y) \rangle_R \propto \delta(\x-\y)$.   By our previous dimension counting argument, $m(\x)$ behaves like a dimension $D/2$ operator.  Similarly, $\psi$ is a dimension $ D/2$ field.  As a result, the stochastic mass term has dimension $\tfrac{3}{2} D$, which is relevant for $D< 2$.   The reason that localization occurs for arbitrarily weak disorder is that its effective strength grows as we look at longer distances.  Any finite amount of disorder will effectively become strong eventually.

It is tempting to draw an analogy between Anderson localization and inflation in many dimensions~\cite{Tye:2007ja,Podolsky:2008du}.  In some such models (see e.g.~\cite{Baumann:2014nda}), inflation is described by the motion of a space-filling brane through higher dimensions.  The brane acts like a particle responding to the forces of a number of localized sources.  From a four-dimensional perspective, the number of extra-dimensions determines the number of scalar fields needed to describe the brane's position.  If there were some analogy of localization, it would prevent inflation from occurring with too few fields by confining the brane to a point.  However, this is not what we found.  

While our analysis is essentially no different than the case of electrons in a metal, there are two important aspects that are responsible for the different conclusion.  First, the disorder potential in inflation involves only time, no matter how many fields play a role during inflation.  The reason this occurs from the higher dimensional perspective is that the brane fills our three spatial dimensions so quantum mechanical effects are suppressed by the volume of space through an effective $\hbar_{\rm effective} = \hbar / V_{\rm 3D} \to 0$.  As a result, the wave functions are just classical trajectories, with no analogue of the interference effects that are important to Anderson localization.  

Second, inflation does not probe the system at arbitrary low energies, but is instead controlled by physics at the Hubble scale.  In this sense, the only question of interest is whether disorder can be perturbatively small at the scale $H$.  As we saw previously, by appropriate choices of parameters, disorder can be perturbative whether it arises from an irrelevant, marginal or relevant operator.  For solids, we are instead interested in the behavior at arbitrarily low energies after fixing a non-zero amplitude for disorder.  This slight difference in the order of limits is largely responsible for the difference between inflation and a solid.

\section{Derivation of the Optimal Estimator}\label{app:estimator}

Given a field $\kappa$ that modulates the temperature fluctuations, the optimal estimator is given by~\cite{Hanson:2009gu}
\beq
\hat \kappa= {\cal F}^{-1} [ \bar \kappa -\langle \bar \kappa\rangle ] \ ,
\eeq
where
\beq 
\bar \kappa = \frac{1}{2} \sum_{\ell,m,\ell', m'} \bar \Theta_{\ell, m}^{\dagger} \partial_{\kappa} C_{\ell, m ; \ell',m'} \bar \Theta_{\ell',m'} \quad , \qquad {\bf \bar \Theta }\equiv {\bf  C}^{-1}_{\kappa=0} \, {\bf \Theta}
\eeq
and
\beq
{\cal F} = \langle \bar \kappa \bar \kappa \rangle - \langle \bar \kappa \rangle \langle \bar \kappa \rangle \ .
\eeq
Here $\Theta = \frac{\delta T}{T}$ is the temperature fluctuation of the CMB and $C_{\ell, m ; \ell',m'}$ is its covariance matrix.  

As we derived in the main text, the observational consequences of disorder on the power spectrum can be recast in the form
\beq
C_{\ell,m; \ell',m'} = C_\ell  \delta_{\ell \ell'} \delta_{m m'} (1+ \kappa_\ell)  + C^{N N}_{\ell,m; \ell',m'}
\eeq
where $C^{N N}_{\ell,m; \ell',m'}$ is the contribution from instrumental noise.  This is sufficient to define the estimator in general using 
\beq
\bar \kappa_\ell = \tfrac{1}{2} \sum_m\bar \Theta^*_{\ell,m} C_\ell \bar \Theta_{\ell,m} \ .
\eeq
This result is most intuitive for cosmic variance limited modes, in which case
\beq
\bar \kappa_\ell = \tfrac{1}{2} (C_\ell)^{-1} \sum_m \Theta^*_{\ell,m} \Theta_{\ell,m}
\eeq
and
\beq
{\cal F}_{\ell, \ell'} = \delta_{\ell, \ell'} \tfrac{1}{2} (2 \ell + 1) + {\cal O} (\kappa_\ell) \ .
\eeq
Combining these results we find
\beq
\hat \kappa_\ell = \frac{1}{(2\ell +1) C_\ell} \Big (\sum_m \Theta^*_{\ell,m} \Theta_{\ell,m} \Big)- 1 = \frac{1}{C_\ell}  \Big (C_\ell^{\rm obs.}-C_{\ell} \Big) \ ,
\eeq
where $\frac{1}{2 \ell +1}\sum_m \Theta^*_{\ell,m} \Theta_{\ell,m} \equiv C_\ell^{\rm obs.}$.

\newpage
\addcontentsline{toc}{section}{References}
\bibliographystyle{utphys}
\bibliography{disorder,books}
\end{document}